\documentclass[twocolumn,showpacs,preprintnumbers,amsmath,amssymb]{revtex4}

\usepackage{graphicx}
\usepackage{dcolumn}
\usepackage{bm}
\usepackage{pstricks}
\usepackage{epsfig}

\begin{document}


\title{Entanglement and quantum interference}

\author{Paul O'Hara}
\affiliation{%
Dept. of Mathematics\\
Northeastern Illinois University
}%



\begin{abstract}
In the history of quantum mechanics, much has been written about the
double-slit experiment, and much debate as to its interpretation has
ensued. Indeed, to explain the interference patterns for sub-atomic
particles, explanations have been given not only in terms of the
principle of complementarity and wave-particle duality but also in
terms of quantum consciousness and parallel universes. In this
paper, the topic will be discussed from the perspective of
spin-coupling in the hope of further clarification. We will also
suggest that this explanation allows for a realist interpretation of
the Afshar Experiment.
\end{abstract}

\pacs{3.65, 3.75-b}
\maketitle

\subsection{\label{sec:level1}Description of the experiment}

\begin{figure}
\centerline{\psfig{figure=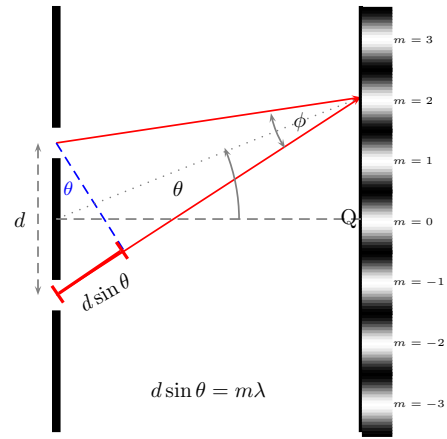,width=2.25in}}
\caption{Interference fringes} \label{spectrum}
\end{figure}

In itself, the double-slit experiment is easy to describe. Let us
assume without loss of generality that we are using electrons to
conduct the experiment. First a beam of electrons is fired through a
single slit, and a normal diffraction pattern observed. Next the
beam of electrons is fired at a pair of slits and an interference
pattern is observed (Fig. \ref{spectrum}). This interference pattern
presents an immediate difficulty. If the electrons were truly
particles then we would expect each electron to have passed through
either one of the slits (but not both) according to the laws of
independent probability.

Specifically, if the particle state is
$\left|\psi\right>=\left|\psi_1\right>+\left|\psi_2\right>$ and if
the electron beam were composed of independent non-interacting
particles then the probability distribution would be given by
$|\psi_1|^2+|\psi_2|^2$. Instead, a distribution of the form
$|\psi_1+\psi_2|^2$ is observed, consistent with a wave-front. This
same interference pattern also emerges if electrons pass through the
slits and hit the screen one at a time, thus excluding the
possibility that they have interacted or interfered with each other.
It would seem individual electrons interfere with themselves.
Moreover, the matter is further complicated when we attempt to
observe which route through the slits the individual electrons might
have taken. To our surprise we will find that the distribution
associated with particles now re-emerges. And so we might wonder,
how did the electron know it was being observed and change back to
behaving like a particle?

Some have tried to explain this in terms of quantum consciousness,
whereby the electron is neither a particle nor a wave, but responds
according to the type of question we ask, and the type of probing we
do. Others would say that it is both a particle and a wave, and that
the projection postulate forces it to choose one or the other state,
according to how we measure it. Again others have attempted to
explain the result in terms of parallel universes or from Cramer's
transactional viewpoint \cite {cr}.

Before attempting to agree or disagree with any of the above
interpretations, I would like to first re-examine the problem
mathematically from the perspective of entanglement and rotational
invariance, and see what this analysis suggests.

\subsection {Rotationally-invariant states}

Part of the difficulty with a strict particle interpretation of
quantum interference is the assumption that particles independently
pass through either one of the slits, with the slits serving merely
as a filter or a conduit for the particles. In this paper, I would
like to suggest that regardless of which path the electron takes, if
the slits are sufficiently close together, then the acquired
spin-state of an individual electron depends upon its interaction
with both of the slits, unless inhibited from doing so. At the same
time this does not mean that the particle has traveled through both
slits, nor does it preclude that the particle has taken a definite
route through the interferometer. However, it does require that we
focus on the spin-state dependency as determined by the two
apertures. Moreover, I think these results will be found not only to
be in full agreement with the quantum eraser effect obtained through
parametric down conversion\cite{or, wal}, but will also highlight
the role of entanglement in interference phenomena, as well as
clarify the role of conditional probability in physics.

\subsubsection{\label{sec:level1}A coupling principle:}

It has been shown in previous work that isotropically
spin-correlated states must occur in pairs \cite {ohara}.
Specifically, if we denote spin-up and spin-down states by
$\left|+\right>$ and $\left|-\right>$, this means that there exist
unique rotationally invariant pure states of the form
\begin{eqnarray} \left|\psi\right>&=&\frac{1}{\sqrt 2}(\left|+\right>\left|-\right>-\left|-\right>\left|+\right>)\\
\left|\psi\right>&=&\frac{1}
{\sqrt2}(\left|+\right>\left|+\right>+\left|-\right>\left|-\right>),
\end{eqnarray}
which correspond to the singlet state and its spinor conjugate
respectively. These entangled states have the characteristic that if
a spin-measurement is made in an arbitrary direction on one of the
states, then the state of the other particle is known with
certainty. For example, in the case of the singlet state, if a
positive spin is detected along an arbitrary z-axis of particle one
then a negative spin value will be detected along the corresponding
z-axis of particle two. Indeed, I would suggest that it is precisely
this spinor correlation which is the primary cause of the
interference phenomena, and if this correlation is broken as often
happens when we measure or try to detect a particle, then the
interference phenomena disappears.

Specifically, let the initial state
\begin{eqnarray} \left|\psi\right>&=&\frac{1}{\sqrt 2}(\left|+\right>\left|+\right>+\left|-\right>\left|-\right>)
\end{eqnarray}
represent the state as registered with respect to the normal at Q
(m=0 in Fig. \ref{spectrum}). In other words, from the perspective
of Q, we not only assume that the incoming particle has come with
equal probability from either aperture, but that its spinor
component along the normal to the screen at Q has been induced by
the apertures working in tandem, as captured in the tensor product
of equation (3). Next, shift the normal at Q to P. In this case, the
pure initial state will transform under rotation operators
$(\mathcal{R(\alpha}),\mathcal{R(\beta)})$ into the mixed quantum
state
\begin{eqnarray} \left|\psi\right>&=&1/{\sqrt
2}[\sqrt{1-\rho^2}(\left|+\right>\left|-\right>-\left|-\right>\left|+\right>)\nonumber\\
&\ &\qquad\qquad+
\rho(\left|+\right>\left|+\right>+\left|-\right>\left|-\right>)]
\end{eqnarray}
where $\rho^2\le 1$, and $\alpha$ and $\beta$ are the angles of
incidence of the straight-line trajectories drawn from the apertures
(Fig.2). Note, this mixed state is composed of two isotropically
spin-correlated pure states describing the spin state of a single
particle arriving at $P$ with probability $\rho(\phi)$, where
$\phi=\alpha-\beta$ is the angle subtended by the two rays
connecting $P$ to the two apertures (Fig. 2). This means that
regardless of which aperture a particle may or may not take, the
relative spin-states are correlated by the slits themselves, in such
a way that the state
$(\left|+\right>\left|+\right>+\left|-\right>\left|-\right>)$ will
be detected with probability $\rho^2$ and failed to be detected with
probability $1-\rho^2$. In fact, failure to detect means that the
other state
$(\left|+\right>\left|-\right>-\left|-\right>\left|+\right>)$ will
have occurred. Also we note that intuitively, regardless of the
particles trajectory, we can by analogy imagine the direction of the
particle's spin behaving like a compass needle, and pointing always
in the direction of the normal to the screen.\newline
\begin{pspicture}(-1,-2.5)(5,3.5)
\psline[linecolor=red,arrows=->](0,-1.25)(5,2)
\psline[linecolor=red,arrows=->](0, 1.25)(5,2)
\psline[linecolor=black,linewidth=4pt](0,-1.9)(0,-1.5)
\psline[linecolor=black,linewidth=4pt](0,-1)(0,1)
\psline[linecolor=black,linewidth=4pt](0,2.5)(0,1.5)
\psline[linecolor=gray,linestyle=dashed,arrows=<->](-0.3,-1.25)(-0.3,1.25)
\rput(-0.6,0){$d$}
\psline[linecolor=black,linewidth=2pt](5,-1.9)(5,2.5)
\psline[linecolor=gray,linestyle=dashed](0,2)(5,2) \rput(5.2,2){$P$}
\psarc[arrows=->,linecolor=gray](5,2){-2.5}{0}{9}\rput(2.8,1.8){$\alpha$}
\psarc[arrows=->,linecolor=gray](5,2){-1.5}{0}{31}\rput(4,1.6){$\beta$}
\rput(2.5,-1.9){{\small Fig. 2}}%
\end{pspicture}

Finally to conclude this section, an explicit derivation of equation
(4) is given, using the SU(2) properties associated with rotational
invariance. Indeed, by our initial assumption, if $\phi=0$ then the
state at $m=0$ is of the form
$\left|+\right>\left|+\right>+\left|-\right>\left|-\right>$ with
intensity $I_0$. Applying the rotation operator
$\mathcal{R(\alpha)}$ to the first component, gives the
corresponding intensity at $P$, resulting from a particle moving
along a ray making an angle of incidence $\alpha$ (Fig. 2) with the
normal. It follows that
\begin{eqnarray}&&(\mathcal{R(\alpha)}\left|+\right>\left|+\right>+\mathcal{R(\alpha)}\left|-\right>\left|-\right>)
\nonumber\\
=&&\left(\begin{array}{cc}
                           \cos(\alpha) & \sin(\alpha) \\
                          -\sin(\alpha) & \cos(\alpha) \\
                         \end{array}
                       \right)
                       \left(\begin{array}{c}
                    1 \\
                    0 \\
                  \end{array}
                \right)
                \otimes
                       \left(
                         \begin{array}{c}
                           1 \\
                           0 \\
                         \end{array}
                       \right)\nonumber\\
                       &&\qquad\qquad+
                       \left(\begin{array}{cc}
                           \cos(\alpha) & \sin(\alpha) \\
                          -\sin(\alpha) & \cos(\alpha) \\
                         \end{array}
                       \right)
                       \left(\begin{array}{c}
                    0 \\
                    1 \\
                  \end{array}
                \right)
                \otimes
                       \left(
                         \begin{array}{c}
                           0 \\
                           1 \\
                         \end{array}
                       \right)
\nonumber\\=&&\cos(\alpha)\left[\left(
         \begin{array}{c}
           1 \\
           0 \\
         \end{array}
            \right)\left(
                \begin{array}{c}
                  1 \\
                  0 \\
                \end{array}
              \right)+\left(\begin{array}{c}
                                 0 \\
                                 1 \\
                               \end{array}
                             \right)\left(
                                      \begin{array}{c}
                                        0 \\
                                        1 \\
                                      \end{array}
                                    \right)\right]
\nonumber\\
&&\qquad -\sin(\alpha)\left[\left(
         \begin{array}{c}
           1 \\
           0 \\
         \end{array}
       \right)\left(
                \begin{array}{c}
                  0 \\
                  1 \\
                \end{array}
              \right)-\left(
                        \begin{array}{c}
                                 0 \\
                                 1 \\
                               \end{array}
                             \right)\left(
                                      \begin{array}{c}
                                        1 \\
                                        0 \\
                                      \end{array}
                                    \right)\right].
                                    \end{eqnarray}
Similarly, if we multiply the first tensor component by the rotation
operator $\mathcal{R(\alpha)}$ and the second component by the
operator $\mathcal{R(\beta)}$, and denote $\alpha-\beta=\phi$, we
obtain:
\begin{eqnarray*}(\mathcal{R(\alpha)},\mathcal{R(\beta)})(\left|+\right>\left|+\right>+\left|-\right>\left|-\right>)=\qquad\qquad\qquad\qquad\qquad
\qquad \qquad \quad &&\nonumber\\
\left(\begin{array}{cc}
                           \cos(\alpha) & \sin(\alpha) \\
                          -\sin(\alpha) & \cos(\alpha) \\
                         \end{array}\right)
                         \left(
                  \begin{array}{c}
                    1 \\
                    0 \\
                  \end{array}
                \right)
                \otimes\left(
                         \begin{array}{cc}
                           \cos(\beta) & \sin(\beta) \\
                          -\sin(\beta) & \cos(\beta) \\
                          \end{array}\right)
                          \left(
                          \begin{array}{c}
                          1\\
                          0\\
                          \end{array}\right)\qquad \qquad &&\\
                +\left(
                         \begin{array}{cc}
                           \cos(\alpha) & \sin(\alpha) \\
                          -\sin(\alpha) & \cos(\alpha) \\
                         \end{array}
                       \right)
                       \left(
                         \begin{array}{c}
                           0 \\
                           1 \\
                         \end{array}
                       \right)
                       \otimes
                       \left(
                         \begin{array}{cc}
                           \cos(\beta) & \sin(\beta) \\
                          -\sin(\beta) & \cos(\beta) \\
                         \end{array}
                       \right)
                       \left(
                         \begin{array}{c}
                           0 \\
                           1 \\
                         \end{array}
                       \right)\qquad \qquad &&
\end{eqnarray*}
which on multiplying out gives
\begin{eqnarray*}
&&\cos(\phi)\left[\left(
         \begin{array}{c}
           1 \\
           0 \\
         \end{array}
            \right)\left(
                \begin{array}{c}
                  1 \\
                  0 \\
                \end{array}
              \right)+\left(
              \begin{array}{c}
               0 \\
               1 \\
               \end{array}
               \right)\left(
               \begin{array}{c}
               0 \\
               1 \\
               \end{array}
               \right)\right]\\
&&\qquad -\sin(\phi)\left[\left(
         \begin{array}{c}
           1 \\
           0 \\
         \end{array}
       \right)\left(
                \begin{array}{c}
                  0 \\
                  1 \\
                \end{array}
              \right)-\left(
                        \begin{array}{c}
                                 0 \\
                                 1 \\
                               \end{array}
                             \right)\left(
                                      \begin{array}{c}
                                        1 \\
                                        0 \\
                                      \end{array}
                                    \right)\right].
\end{eqnarray*}
This is identical to equation (4) above, and further justifies our
interpretation. Also note that $\phi$ and $\theta$ are related by
the equation $\phi=\frac{2\pi d}{\lambda}\sin(\theta)$
(Fig.\ref{spectrum}) \cite{hr}.

\subsubsection{\label{sec:level2}The detection process:}

We now ask what happens when we try to detect the particle route?
Mathematically, the answer is quite straight forward and can be
handled using projection operators. Specifically, the transmitted
rotationally invariant state, given by
$\left|\psi\right>=1/\sqrt{2}(\left|+\right>\left|+\right>+\left|-\right>\left|-\right>),$
is reduced by the detector (placed at slits 1 and/or 2) to the state
\begin{eqnarray*}P_i
\left|\psi\right>&=&1/\sqrt{2}P_i(\left|+\right>\left|+\right>+\left|-\right>\left|-\right>)\\
&=&1/\sqrt{2}(\left|+\right>_i+\left|-\right>_i) \end{eqnarray*}
where $i=1$ or 2 represents the aperture from which the particle is
emitted. In other words, from the perspective of this paper, the
measuring process breaks the entanglement, causing the interference
to disappear. The spins are no longer correlated and the probability
distribution is now associated with two independent random variables
representing the proportion of particles emerging respectively from
each aperture. In neither case, does it follow that the particle has
entered both slits simultaneously and interacted with itself.
Rather, the double slit experiment seems to be telling us that
regardless of the slit from which the particle leaves, both slits
have contributed to inducing an isotropically-polarized state
relative to the slits themselves, which is then manifested in an
interference effect, unless interrupted by the detection process.
Moreover, this perspective is not only compatible with the Afshar
experiment, but also offers a realist interpretation of his results
\cite{afs}.

Finally, it should be noted that if the above interpretation is
correct, then it should also be possible to design an experiment to
destroy the correlation using Stern-Gerlach magnets, without ever
obtaining ``which way'' information; thereby demonstrating that it
is not our knowledge of the situation that determines the final
quantum state, but rather the way in which the states are correlated
and transformed from coherent states into non-coherent states.  An
experiment similar to this is described by Qureshi and Rehman in
\cite{qr}. However, in their case they first use a magnet to obtain
``which way'' information as the particles enter the slit and then
erase this information with another magnet after the particles exit
the slits, to produce interference. What is being proposed here is
the opposite. First particles enter the slits undetected and form an
interference pattern on the screen. Then a Stern-Garlach magnet
field, placed orthogonal to the screen and its normal, is turned on
as the particles emerge from the slits. If the above theory is
correct then the interference pattern will disappear by
strategically positioning the magnet.
\newline
\begin{pspicture}(-0.5,-3.5)(.5,4.0)
\psline[linecolor=red,arrows=->](0,-1.25)(5,2)
\psline[linecolor=red,arrows=->](0, 1.25)(5,2)
\psline[linecolor=red,linestyle=dashed](0,0)(5,2)
\psline[linecolor=black,linewidth=4pt](0,-3.5)(0,-1.5)
\psline[linecolor=black,linewidth=4pt](0,-1)(0,-.25)
\psline[linecolor=black,linewidth=4pt](0,.25)(0,1)
\psline[linecolor=black,linewidth=4pt](0,1.5)(0,3.5)
\psline[linecolor=black,linewidth=4pt](0,3.5)(0,1.5)
\psline[linecolor=gray,linestyle=dashed,arrows=<->](-0.3,-1.25)(-0.3,1.25)
\rput(-0.6,0){$d$} \rput(0,-1.25){$a_3$} \rput(0,0){$a_2$}
\rput(0,1.25){$a_1$}
\psline[linecolor=black,linewidth=2pt](5,-3.5)(5,3.5)
\psline[linecolor=gray,linestyle=dashed](0,0)(5,0)
\psline[linecolor=gray,linestyle=dashed](0,2)(5,2)
\psarc[arrows=<->,linecolor=gray](5,2){-3.0}{23}{33}\rput(2.5,1.4){$\phi_{12}$}
\psarc[arrows=<->,linecolor=gray](5,2){-3.0}{10}{20}\rput(2.65,.75){$\phi_{23}$}
\psarc[arrows=<->,linecolor=gray](5,2){-1.5}{10}{33}\rput(4,1.6){$\phi_{13}$}
\rput (5.2,2){P}%
\rput(2.5,-3.3){\small Fig. 3}%
\end{pspicture}

\subsubsection{\label{sec:level3}Multiple slits:}

In this section, we briefly analyze the interference phenomena
caused by three or more slits. At first this might seem a formidable
problem to resolve, given the many ways that the slits may influence
the polarization. However, because of the coupling principle, we
need only worry about paired states.

For example, in the case of three slits lying on a line, the
possible paired states are of the form, $\left|\psi_{12}\right>$,
$\left|\psi_{23}\right>$ or $ \left|\psi_{13}\right>$, where each
$\left|\psi_{ij}\right>$ equals
$$\cos(\phi_{ij})(\left|+\right>\left|+\right>+\left|-\right>\left|-\right>)+
\sin(\phi_{ij})(\left|+\right>\left|-\right>+\left|-\right>\left|+\right>)$$
with $\phi_{ij}$ being the angle subtended by rays drawn from $a_i$
and $a_j$ and meeting at P (Fig. 3). Moreover, it is not difficult
to show that once we have the correct correlation written down for
any pair of slits, then it is easy to deduce the correlation
associated with any other pair of slits. Indeed, it seems remarkable
that multiple slit correlations can be reduced to paired
correlations. To show this explicitly it is sufficient to consider
an interferometer made of three apertures:\newline Let
$\textbf{u}=\frac{1}{\sqrt 2}
\left|+\right>\left|+\right>+\left|-\right>\left|-\right>$ and
$\textbf{v}=\frac{1}{\sqrt 2}
\left|+\right>\left|-\right>-\left|-\right>\left|+\right>$ then
equation (4) reduces in this notation to the compact form:
\begin{eqnarray}[\mathcal{R(\alpha),R(\beta)}]\textbf{u}&=&\cos(\beta-\alpha)\textbf{u}-\sin(\beta-\alpha)\textbf{v}\nonumber\\
&=&[\mathcal{I, R(\beta-\alpha)}]\textbf{u}. \end{eqnarray}
Similarly, we can show by direct calculation that
\begin{eqnarray}[\mathcal{R(\alpha),R(\beta)}]\textbf{v}&=&\sin(\beta-\alpha)\textbf{u}+\cos(\beta-\alpha)\textbf{v}\nonumber\\
&=&[\mathcal{I, R(\beta-\alpha)}]\textbf{v}. \end{eqnarray}
Combining these two results and recalling that
$\left|\psi_{12}\right>=[\mathcal{I, R(\beta-\alpha)}]\textbf{u}$
allows us to write
\begin{eqnarray}[\mathcal{R(\beta),R(\gamma)}]\left|\psi_{12}\right>&=&
[\mathcal{I,R(\gamma-\beta)}]\left|\psi_{12}\right>\nonumber\\
&=&[\mathcal{I, R(\gamma-\beta)}]
[\mathcal{I,R(\beta-\alpha)}]\textbf{u}\nonumber\\
&=&[\mathcal{I, R(\gamma-\beta+\beta-\alpha)}]\textbf{u}\nonumber\\
&=&[\mathcal{I, R(\gamma-\alpha)}]\textbf{u}\nonumber\\
&=&\cos(\gamma-\alpha)\textbf{u}-\sin(\gamma-\alpha)\textbf{v}\nonumber\\
&=&\cos(\phi_{13})\textbf{u}-\sin(\phi_{13})\textbf{v}\nonumber\\
&=&\left|\psi_{13}\right>
\end{eqnarray}

There is a certain beauty in this result. It shows that if at any
point $P$ we can write down the state function for any pair of
apertures, then by means of rotation operators we can also transform
it into the corresponding state function for any other pair. To do
so, it is sufficient to know the angle a ray from the aperture makes
with the normal to point $P$. For example, if we were to modify the
above and work with rotations $(\mathcal{R(\alpha),R(\beta)})$ and
$(\mathcal{R(\alpha),R(\gamma)})$ we could have obtained the state
function $\left|\psi_{23}\right>$.
\newline

\noindent Note: (1) In the case of spin 0 particles like mesons the
above results would also apply, provided some type of isotropic
polarized states occur. A photon may be used as an example of this
type of polarization.\newline (2) In this paper we have assumed that
$\left|+\right>\left|+\right>+\left|-\right>\left|-\right>$ is
transmitted and the state
$\left|+\right>\left|-\right>-\left|-\right>\left|+\right>$ is
absorbed. However, the theory is completely symmetrical and we could
equally have assumed the opposite without any loss of generality.
Indeed, in the case of the quantum eraser effect, it could be argued
that in some cases the state
$\left|+\right>\left|-\right>-\left|-\right>\left|+\right>$ is
registered on the detector, while the,
$\left|+\right>\left|+\right>+\left|-\right>\left|-\right>$ is
absorbed and not detected.

\subsection {Conclusion}

By associating quantum interference with rotationally-invariant
states induced by the slits, we have succeeded in showing that it is
possible for an object such as an electron or photon to exhibit both
particle and wave properties in a non-mutually exclusive way.
Furthermore, from the perspective of this paper, there is no
contradiction in considering these objects to be particles at all
times, in that they take a definite route through the
interferometer, while simultaneously exhibiting wave characteristics
associated with the isotropically-spin correlated states and induced
by the apertures themselves, in accordance with the laws of
conditional probability. Indeed, to the extent that we consider the
object to be a particle then it will always be a particle. However
to the extent that the spin (or polarization) states are defined and
correlated relative to the apertures, they will also exhibit wave
characteristics. Nevertheless, to capture these properties
simultaneously is very difficult in that the act of path detection
is also responsible for undoing the spin-correlation, which also
means to undo the wave characteristics. Wave properties need to be
detected prior to detecting the particle properties if both
properties are to be exhibited. The two observables do not commute,
but at the same time, it seems to me that the Afshar experiment has
succeeded in demonstrating that these two properties can co-exist
simultaneously, in a way that is compatible with a realist
interpretation of complementarity.

Finally, it is worth emphasizing that if the above formulism were to
be regarded as correct, then in my opinion the focus of
interferometer research should move away from particle detection to
trying to better understand how the correlated states are induced.

\bibliography{apssamp}

\end{document}